\documentclass[epj]{svjour}
\usepackage{latexsym}
\usepackage{graphics}
\begin{document}
\title{$Q\bar{Q}$ modes in the Quark-Gluon Plasma}
\author{D.~Cabrera \and R.~Rapp
}                     
%
%
\institute{Cyclotron Institute and Physics Department, Texas A\&M University,
College Station, Texas 77843-3366, U.S.A.}
\date{Received: date / Revised version: date}
%
\abstract{
We study the evolution of heavy quarkonium states with temperature in a
Quark-Gluon Plasma by evaluating an in-medium $Q\bar{Q}$ $T-$matrix 
within a reduced Bethe-Salpeter equation in $S-$ and $P-$wave channels. 
The interaction kernel is extracted from finite-temperature QCD lattice 
calculations of the singlet free energy of a $Q\bar{Q}$ pair.   
Quarkonium bound states are found to gradually move across the $Q\bar{Q}$
threshold after which they rapidly dissolve in the hot system.
We calculate Euclidean-time correlation functions 
and compare to results from lattice QCD. 
We also study finite-width effects in the heavy-quark 
propagators.
} 
\maketitle
\section{Introduction}
\label{sec:intro}

Bound states of heavy (charm and bottom) quarks ($Q$= $b$, $c$) are 
valuable spectroscopic objects in Quantum Chromodynamics 
(QCD)~\cite{Brambilla:2004wf}.
When embedded into hot and/or dense matter a large class of medium 
modifications can be studied, including (Debye-) color-screening of the 
$Q\bar Q$ interaction, dissociation reactions induced by constituents 
of the medium, and the change in thresholds caused by mass (or width) 
modifications of open heavy-flavor states.
Lattice QCD (lQCD) calculations have made substantial progress in 
characterizing in-medium quarkonium properties from first principles. 
In particular, it has been found that ground state 
charmonia~\cite{Asakawa:2003re,Datta:2003ww,Umeda:2002vr}
and bottomonia~\cite{Petrov:2005ej} do not dissolve until significantly
above the critical temperature, $T_c$, which has been supported in model  
calculations based on potentials extracted from lQCD, using either 
a Schr\"odinger equation for the bound-state 
problem~\cite{Shuryak:2003ty,Wong:2004zr,Alberico:2005xw,Mocsy:2005qw},
or a $T$-matrix approach which additionally accounts for scattering 
states~\cite{Mannarelli:2005pz}.
More reliable comparisons to lQCD can be performed using (spacelike) 
Euclidean-time correlation functions~\cite{Rapp:2002pn,Mocsy:2004bv}, 
which are readily evaluated in lQCD.
The conversion of (timelike) model spectral functions 
requires a description not only of the bound state part of the
spectrum but also its continuum and threshold properties.
 
In the present work we evaluate Euclidean correlation functions
for charmonium and bottomonium in a $T-$matrix approach. 
The basic input consists of in-medium 
$Q\bar Q$ potentials extracted from lQCD, inserted in
a scattering equation to calculate the in-medium $Q\bar Q$ 
$T-$matrix~\cite{Mannarelli:2005pz}. This incorporates bound and 
scattering states on an equal footing (based on the same interaction), 
and additionally enables a straightforward implementation of
in-medium single-particle (quark) properties via self-energy insertions 
in the two-particle Green's function.

\section{Scattering equation and bound states}
\label{sec:BS}

The $T$-matrix equation for $Q\bar{Q}$ scattering in the center of
mass frame and in partial wave basis reads~\cite{Mannarelli:2005pz},
\begin{eqnarray}
\label{LS}
T_l(E;q',q) &=& V_l(q',q) 
\nonumber \\
&-& \frac{2}{\pi} \int_0^{\infty} dk \, k^2 \, 
V_l(q',k)\, G_{\bar{Q}Q}(E;k) \, T_l(E;k,q) \ ,
\nonumber \\
\end{eqnarray}
which follows from a standard 3-dimensional reduction of the 
Bethe-Salpeter equation~\cite{Blankenbecler:1965gx}. 
$G_{\bar{Q}{Q}}(E;k)$ denotes the intermediate two-particle propagator 
including quark selfenergies ($\Sigma$) and Pauli blocking. The 
$T-$matrix equation (\ref{LS}) is solved with the Haftel-Tabakin 
algorithm~\cite{Haftel:1970}, in which the integral equation 
is solved by discretizing 3-momentum, 
$\sum_{k=1}^N {\mathcal F}(E)_{ik} \,  T(E)_{kj} = V_{ij}$, and 
subsequent matrix inversion. The zeroes of 
$\det {\mathcal F}(E)$ for $E < E_{th}$ determine 
heavy quark-antiquark bound states.

The quark selfenergy, $\Sigma$, 
encodes the interactions with (light) quarks and gluons from the heat
bath~\cite{Mannarelli:2005pz}. In the present work we consider 
a fixed heavy-quark mass $m_Q$ (i.e., $\textrm{Re}\,\Sigma=0$) together 
with a small imaginary part, $\textrm{Im}\,\Sigma=-$0.01~GeV,
for numerical purposes.
Effects of temperature dependent heavy-quark masses 
and widths are investigated in Ref.~\cite{cabrerarapp}. 
%
\begin{figure}
\begin{center}
\resizebox{0.3\textwidth}{!}{%
  \includegraphics{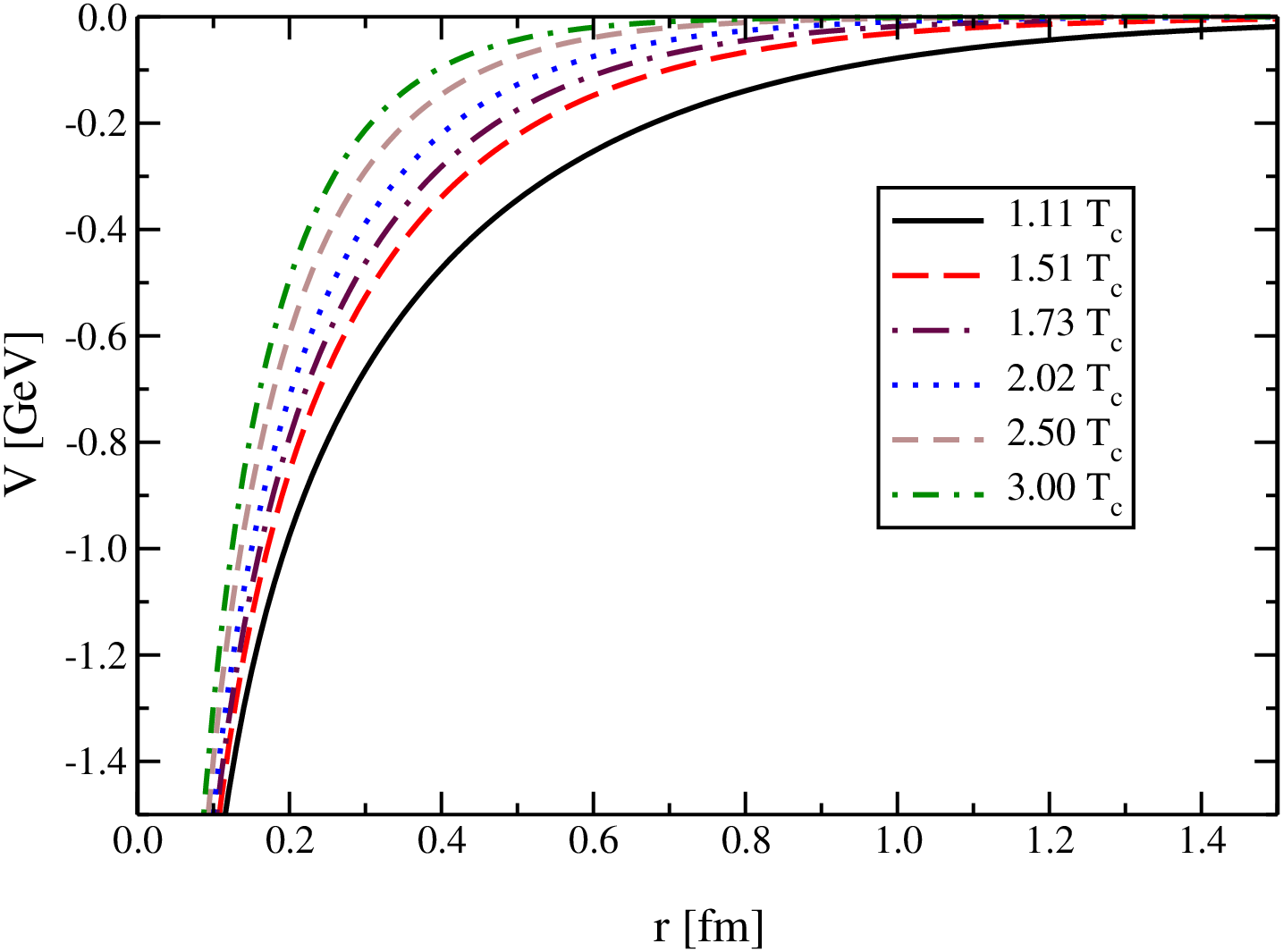}
}
\end{center}
\caption{$Q\bar{Q}$ potential for several temperatures above $T_c$ based on the
color-singlet internal energy.}
\label{fig:V1}       
\end{figure}

The kernel of the scattering equation, $V$, can be estimated from the 
lQCD heavy-quark free energies, $F_1$, even though there is still
an ongoing discussion on how to properly do it. 
Here we identify the heavy-quark potential with the color-singlet 
internal energy, $U_1=F_1 - T \, \frac{dF_1}{dT}$, which reproduces 
ground-state charmonium dissociation temperatures as found in lattice 
analysis of spectral 
functions~\cite{Wong:2004zr,Alberico:2005xw,Mocsy:2005qw,Mannarelli:2005pz}.
In Fig. \ref{fig:V1} we show the $Q\bar{Q}$ potential, $V(r,T) = U_1(r,T) -
U_1(r\to\infty, T)$, as obtained from a fit to the
lQCD color-singlet free energy data~\cite{Petreczky:2004priv}, previously
employed in~\cite{Mannarelli:2005pz}.
The potential evolves smoothly with temperature, decreasing both in magnitude
and range. Different parameterizations of the lattice data
imply sizable uncertainties in the potential through the thermal
derivative of the free energy. These uncertainties are studied in
Ref.~\cite{cabrerarapp} by alternatively obtaining the heavy-quark 
potential from a direct fit to
the lQCD internal energy data of Ref.~\cite{Kaczmarek:2005gi}.
$V_l(q',q)$ follows from a Fourier transform and partial-wave
expansion.

\section{Quarkonium $T-$matrices in the QGP}
\label{sec:T}
We start the calculation of the in-medium $Q\bar Q$ $T-$matrices 
by fixing the heavy-quark masses so that the corresponding quarkonium 
ground states approximately agree with their vacuum masses for the 
lowest considered temperature ($T=1.1T_c$), i.e., 
$m_c = 1.7$~GeV and $m_b = 5.1$~GeV.
%
\begin{figure}
\resizebox{0.45\textwidth}{!}{%
  \includegraphics{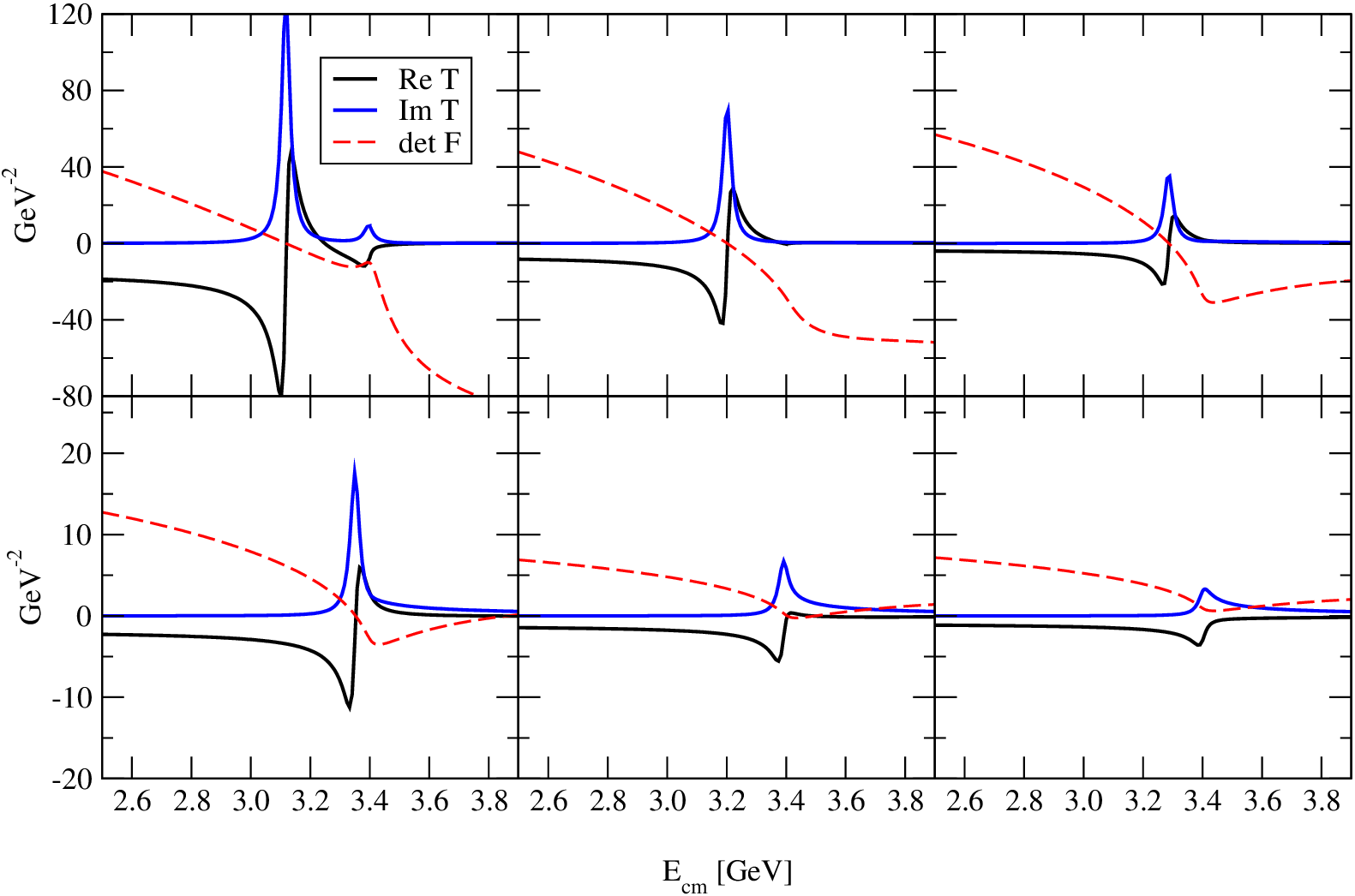}
}
\caption{$T-$matrix for $S-$wave $c\bar{c}$
scattering based on the potential in Fig.~\ref{fig:V1}.
Also shown is $\det {\cal F}$ (dashed line,
arbitrary units). From left to right (up-down) the temperatures 
are $(1.1,1.5,2.0,2.5,3.0,3.3)\,T_c$.}
\label{fig:TccS}       
\end{figure}
Fig.~\ref{fig:TccS} summarizes the on-shell $S-$wave $c\bar{c}$ 
scattering amplitude as a function of the CM energy, for several 
temperatures.
We do not include the hyperfine (spin-spin) interactions and 
therefore $\eta_{c}$ ($\eta_b$) and $J/\psi$ ($\Upsilon$) states are 
degenerate. 
At the lowest temperature, we recover the charmonium ground 
state at $E \approx 3.10$~GeV, whereas the first excited state ($\psi'$) 
has just about melted.
As temperature increases, the $J/\psi(1S)$ gradually 
moves toward threshold and the $T-$matrix is appreciably 
reduced. The bound state survives up to
$\sim3 \, T_c$, where it crosses the $c\bar{c}$ threshold and rapidly
melts in the hot system.

%
\begin{figure}
\resizebox{0.45\textwidth}{!}{%
  \includegraphics{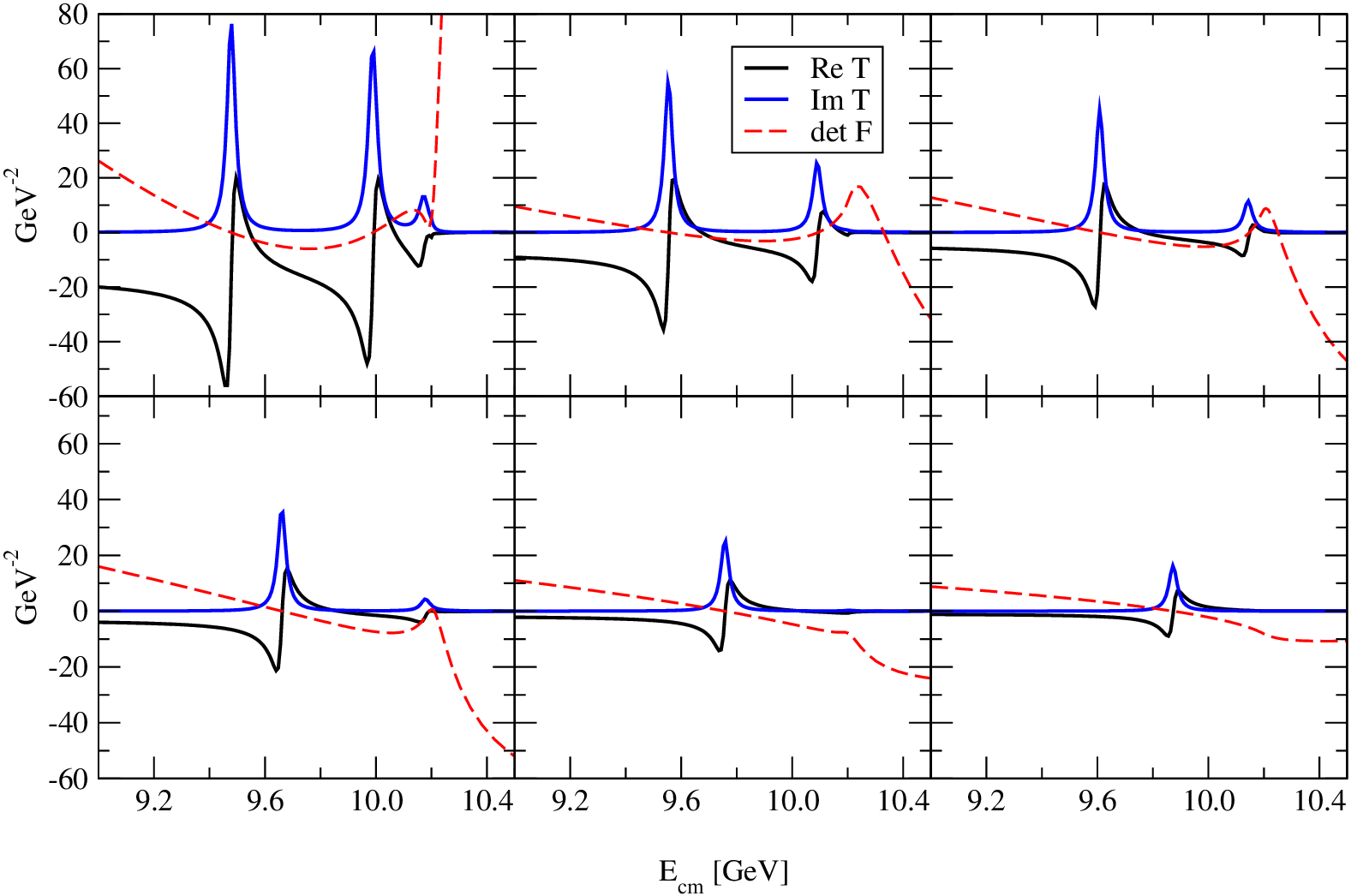}
}
\caption{Same as in Fig.~\ref{fig:TccS} but for $S-$wave $b\bar{b}$
scattering. From left to right and up to down the temperatures are
$(1.1,1.5,1.8,2.1,2.7,3.5)\,T_c$.}
\label{fig:TbbS}       
\end{figure}

The $S-$wave $b\bar{b}$
$T-$matrix exhibits two bound states at the lowest temperature 
($E \approx 9.45,\, 9.95$~GeV for $\Upsilon(1S)$,
$\eta_b$ and $\Upsilon(2S)$, $\eta_b'$, respectively)
and the remnant of a third 
one, which is (almost) melted in the medium, cf.~Fig.~\ref{fig:TbbS}. 
The $\Upsilon(2S)$ 
moves across the $b\bar{b}$ threshold at $T \approx 2.1 \, T_c$, 
whereas the $1S$ state survives in the QGP until much higher 
temperatures, beyond $T \approx 3.5 \, T_c$.

We only find one $P-$wave bound state for the charm system at the lowest
temperature, at $E \approx 3.4$~GeV, which we associate with the $\chi_c$. 
The $P-$wave $b\bar{b}$ system exhibits two bound states at 1.1\,$T_c$, 
with energies $E=9.90,10.15$~GeV, in good agreement with the
nominal values for $\chi_b(1P)$ and $\chi_b(2P)$ in the vacuum.  
The latter moves beyond threshold at $T\approx 1.3\,T_c$, and the 
$(1P)$ state at $T\approx 2.3\,T_c$. Masses and binding energies, 
$(E_B=E_{th}-M)$, of the $P-$wave states are summarized in 
Tab.~\ref{Ptable} for several temperatures.

%
\begin{table}
\caption{Masses and binding energies (in [GeV]) for $P-$wave 
quarkonia as obtained from $\det {\mathcal F}(E)=0$.}
\label{Ptable}       
\begin{tabular}{l|lllll}
\hline\noalign{\smallskip}
$T/T_c$ & 1.1 & 1.3 & 1.5 & 2 & 2.3\\
\noalign{\smallskip}\hline\noalign{\smallskip}
$M[\chi_c(1P)]$ & 3.4 & - & - & - & - \\
$E_B[\chi_c(1P)]$ & $\approx$ 0 & - & - & - & - \\
$M[\chi_b(1P)]$ & 9.90 & 9.98 & 10.04 & 10.14 & 10.20 \\
$E_B[\chi_b(1P)]$ & 0.30 & 0.22 & 0.16 & 0.06 & $\approx$ 0 \\
$M[\chi_b(2P)]$ & 10.15 & 10.20 & - & - & - \\
$E_B[\chi_b(2P)]$ & 0.05 & $\approx$ 0 & - & - & - \\
\noalign{\smallskip}\hline
\end{tabular}
\end{table}

\section{Spectral functions and Euclidean correlators}
\label{sec:SandG}
The Euclidean-time correlation function is defined as the thermal 
two-point mesonic correlation function in a mixed 
Euclidean-time-momentum representation (here $\vec{p}=\vec{0}$), 
\begin{equation}
\label{EuCorr}
G(\tau,T) = \int_0^{\infty} d\omega \, \sigma(\omega,T) \, 
\frac{\cosh[\omega (\tau - \beta/2)]}{\sinh(\omega
\beta/2)} \ ,
\end{equation}
where $\sigma$ is the spectral function obtained from closing
external legs of the in-medium $T-$matrix, schematically as
\begin{equation}
G(E) = \int G_{\bar{Q}Q} + \int G_{\bar{Q}Q}\, T \, G_{\bar{Q}Q}
 \, ,\
\sigma (E) \propto \textrm{Im}\,G(E) \ .
\end{equation}

\begin{figure}
\resizebox{0.45\textwidth}{!}{%
  \includegraphics{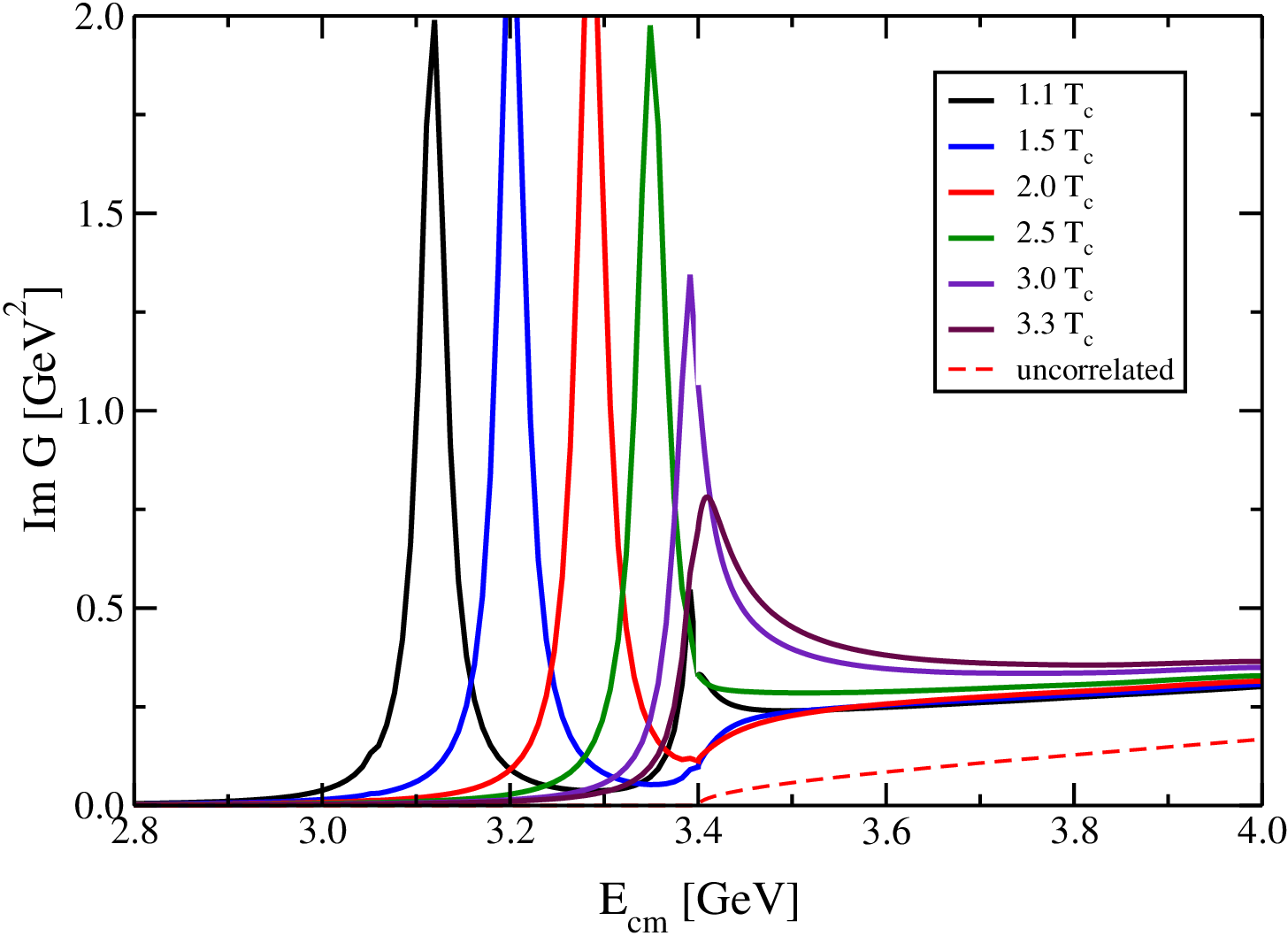}
  \includegraphics{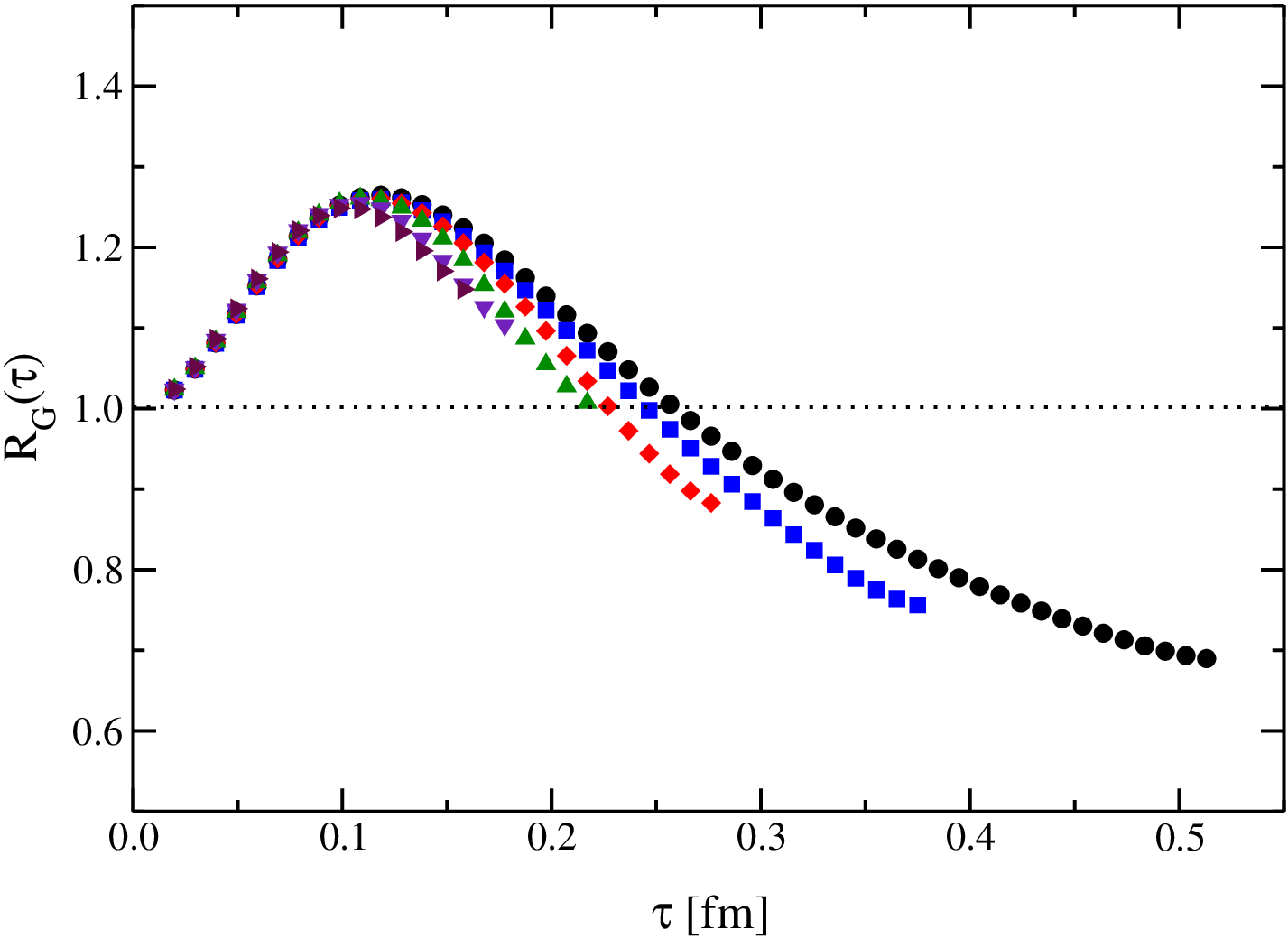}
}
\caption{Left: $c\bar{c}$ spectral
function for $S-$wave scattering at several temperatures. Right: 
normalized correlation function at several temperatures.}
\label{fig:SpecCorr-ccS}
\end{figure}

As expected, the $S-$wave charmonium spectral function  
(left panel of Fig.~\ref{fig:SpecCorr-ccS}) reflects the bound
states of the $T-$matrix, but the (non-perturbative) $c\bar c$ 
rescattering  also generates appreciable strength above 
threshold, where the remnant of first excited state 
($\psi(2S)$) can be still inferred at $T=1.1 Tc$.
The corresponding correlation function (right panel of 
Fig.~\ref{fig:SpecCorr-ccS}) is normalized to a ``reconstructed'' 
correlator represented by a zero-temperature spectral function consisting 
of a $\delta$-function like bound state spectrum and perturbative 
continuum with onset at $E_{cont}=4.5$~GeV~\cite{Mocsy:2005qw}.
The temperature evolution of the correlator is a combined result of 
a decrease in binding energy of the bound states and the contribution 
of the non-perturbative continuum. The sizable drop at large $\tau$ is 
in qualitative agreement with lQCD \cite{Datta:2003ww}. The latter
exhibit somewhat less reduction, leaving room for the effects of 
a threshold reduction with temperature in our $T-$matrix.

\begin{figure}
\resizebox{0.45\textwidth}{!}{%
  \includegraphics{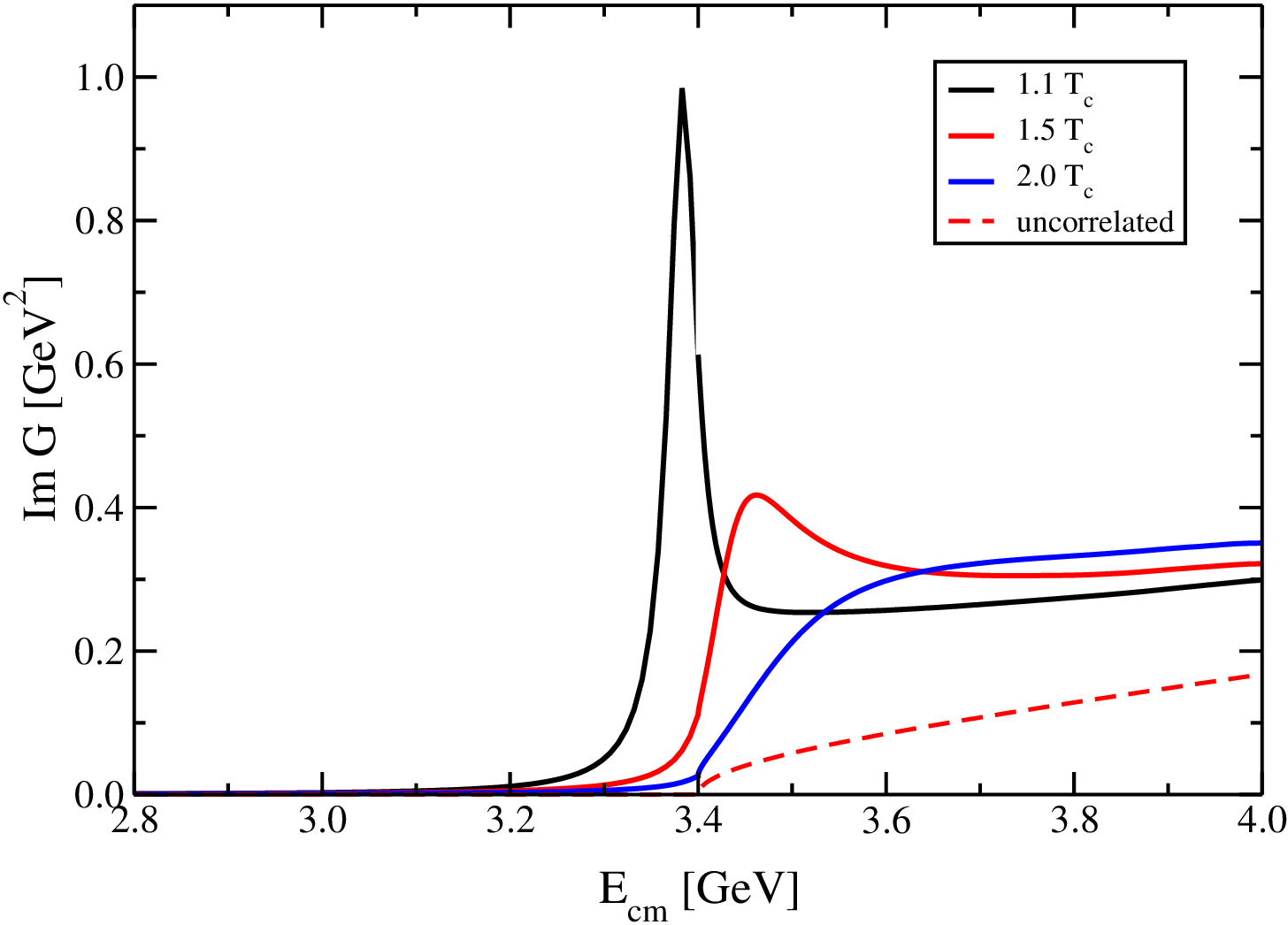}
  \includegraphics{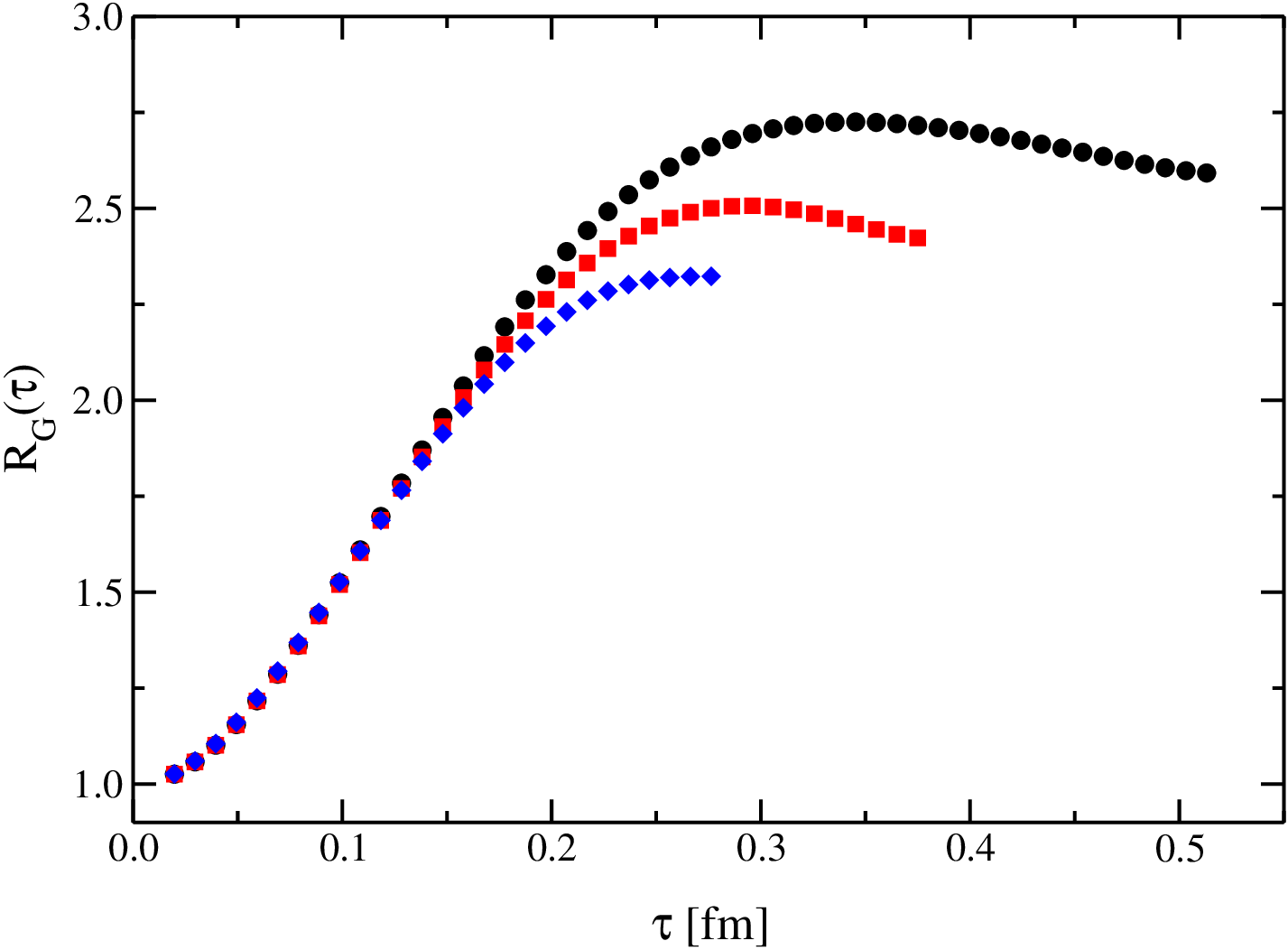}
}
\caption{Same as in Fig. \ref{fig:SpecCorr-ccS} for $c\bar{c}$ $P-$wave
scattering.}
\label{fig:SpecCorr-ccP}
\end{figure}

The $P-$wave charmonium spectral function (Fig.~\ref{fig:SpecCorr-ccP},
left) exhibits one bound state ($\chi_c$) just below the threshold, which 
rapidly melts into the continuum accompanied by a sizable threshold 
enhancement.  The normalized correlator steeply rises in the 
low-$\tau$ regime, due to: (i) non-perturbative rescattering and (ii) 
a larger threshold in the "reconstructed" correlator.
While this is qualitatively consistent with lQCD, the evolution with 
temperature is opposite, which could improve by a shift of strength 
to lower energies due to a decreasing heavy-quark mass with temperature.
The bottomonium correlation functions  
follow a similar pattern as for the 
charmonium system.

\begin{figure}
\begin{center}
\resizebox{0.225\textwidth}{!}{%
  \includegraphics{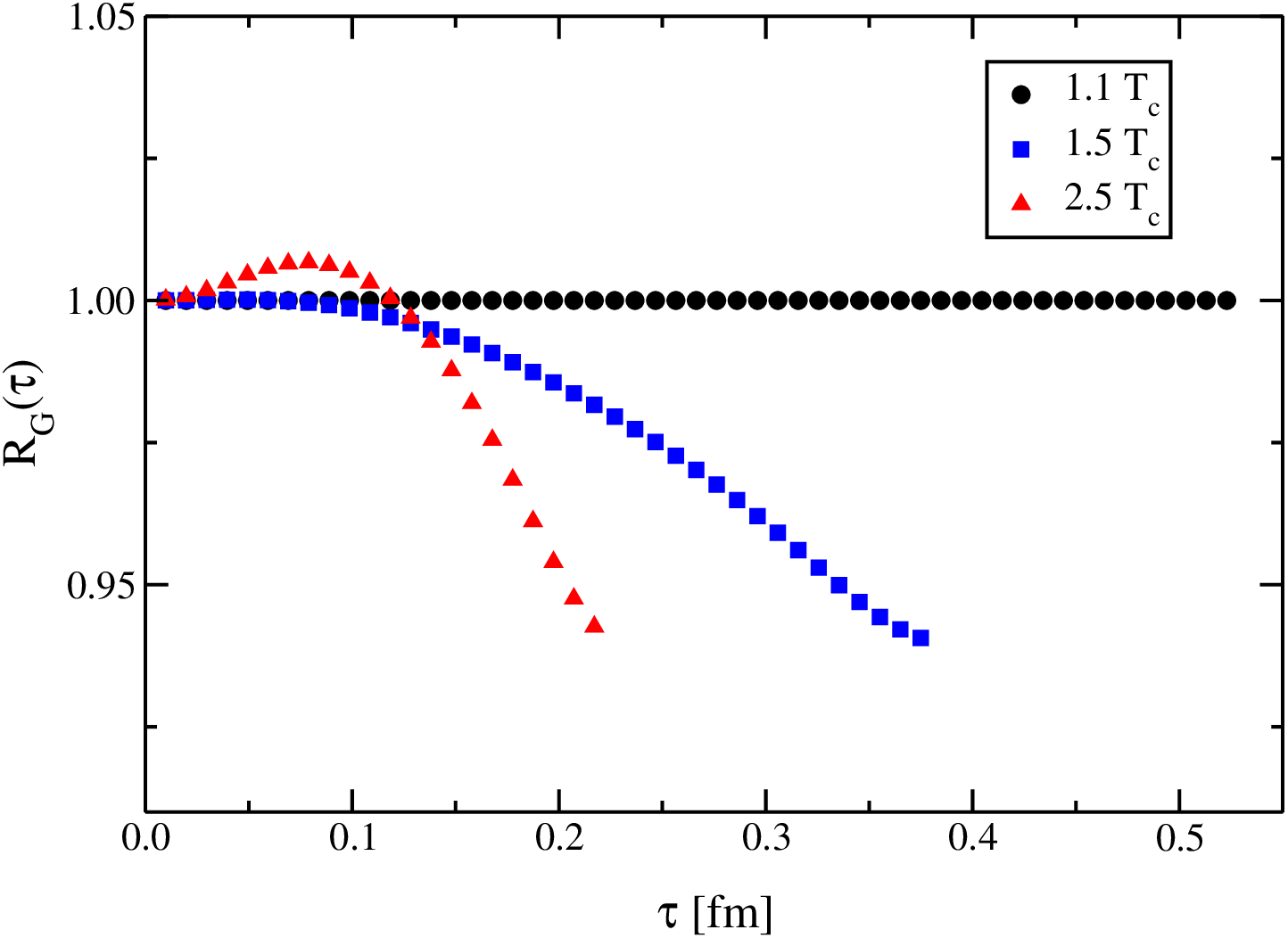}
}
\end{center}
\caption{$S-$wave charmonium correlators when replacing the
``reconstructed" correlator (used for normalization) by the calculated
 result at 1.1\,$T_c$.}
\label{fig:Corr-cc-11Tc}
\end{figure}

It turns out that the $\tau$-dependence of the (normalized) Euclidean
correlators is rather sensitive to the ``reconstructed" correlator used
for normalization. However, $S-$wave charmonium correlators from lQCD 
show rather little temperature dependence below $\sim$1.5\,$T_c$. Thus, 
to reduce ambiguities induced by the reconstructed correlator, we have 
normalized our correlators from the $T-$matrix to the result at 
$T=1.1\,T_c$ calculated in the same approach, 
cf.~Fig.~\ref{fig:Corr-cc-11Tc}. The $\tau$-dependence is now 
substantially weaker than in the right panel of 
Fig.~\ref{fig:SpecCorr-ccS} (limited
to less than 10\%), and the agreement with lQCD~\cite{Datta:2003ww}
is much improved.

We have also studied finite width effects for quarkonia by implementing
$c$-quark widths of 0.05~GeV~\cite{vanHees:2004gq}, inducing
$\Gamma_{\Psi}\approx 0.1$~GeV. The correlators show variations of the 
order of a few percent, indicating a small
sensitivity to phenomenological (sizable) values of quarkonium widths.


%
%

%
%

%

\section{Acknowledgments}
\label{sec:acknow}
We thank M.~Mannarelli and F.~Zantow for providing their (fits to)
lattice QCD results, and M.~Mannarelli and H.~van Hees for
useful discussions.
This work is supported in part by Ministerio de Educaci\'on y
Ciencia (Spain) via a postdoctoral fellowship and by a U.S.~National 
Science Foundation CAREER Award under Grant No. PHY0449489.

%

\begin{thebibliography}{}
%
%






\bibitem{Brambilla:2004wf}
  N.~Brambilla {\it et al.},
  arXiv:hep-ph/0412158.


\bibitem{Asakawa:2003re}
  M.~Asakawa and T.~Hatsuda,
  Phys.\ Rev.\ Lett.\  {\bf 92} (2004) 012001.
  
  
\bibitem{Datta:2003ww}
  S.~Datta, F.~Karsch, P.~Petreczky and I.~Wetzorke,
  Phys.\ Rev.\ D {\bf 69} (2004) 094507.


\bibitem{Umeda:2002vr}
  T.~Umeda, K.~Nomura and H.~Matsufuru,
  Eur.\ Phys.\ J.\ C {\bf 39S1} (2005) 9.

\bibitem{Petrov:2005ej}
  K.~Petrov, A.~Jakovac, P.~Petreczky and A.~Velytsky,
  PoS {\bf LAT2005} (2006) 153.


\bibitem{Shuryak:2003ty}
  E.~V.~Shuryak and I.~Zahed,
  Phys.\ Rev.\ C {\bf 70} (2004) 021901.

\bibitem{Wong:2004zr}
  C.~Y.~Wong,
  Phys.\ Rev.\ C {\bf 72}  (2005) 034906.

\bibitem{Alberico:2005xw}
  W.~M.~Alberico, A.~Beraudo, A.~De Pace and A.~Molinari,
  Phys.\ Rev.\ D {\bf 72} (2005) 114011.

\bibitem{Mocsy:2005qw}
  A.~Mocsy and P.~Petreczky,
  Phys.\ Rev.\ D {\bf 73} (2006) 074007.

\bibitem{Mannarelli:2005pz}
  M.~Mannarelli and R.~Rapp,
  Phys.\ Rev.\ C {\bf 72} (2005) 064905.



\bibitem{Rapp:2002pn}
  R.~Rapp,
  Eur.\ Phys.\ J.\ A {\bf 18} (2003) 459.

\bibitem{Mocsy:2004bv}
  A.~Mocsy and P.~Petreczky,
  Eur.\ Phys.\ J.\ C {\bf 43} (2005) 77.

\bibitem{Blankenbecler:1965gx}
  R.~Blankenbecler and R.~Sugar,
  Phys.\ Rev.\  {\bf 142} (1966) 1051.


\bibitem{Haftel:1970}
  M.~I.~Haftel and F.~Tabakin,
  Nucl.\ Phys.\ A {\bf 158} (1970) 1.

\bibitem{cabrerarapp}
D.~Cabrera and R.~Rapp, in preparation (2006).


\bibitem{Petreczky:2004priv}
P.~Petreczky, private communcation (2004).



\bibitem{Kaczmarek:2005gi}
  O.~Kaczmarek and F.~Zantow,
  arXiv:hep-lat/0506019.




\bibitem{vanHees:2004gq}
  H.~van Hees and R.~Rapp,
  Phys.\ Rev.\ C {\bf 71} (2005) 034907.




\end{thebibliography}
%

\end{document}